\documentclass[12pt,preprint]{aastex}
\usepackage[nonamelimits]{amsmath}
\usepackage{graphicx}
\begin{document}

\def\der#1#2{{\partial#1\over\partial#2}}
\def\derss#1#2{{\partial^2#1\over\partial#2^2}}

\title{Power-law Tails from Dynamical Comptonization \\
in Converging Flows}

\author{Roberto Turolla\altaffilmark{1}, Silvia
Zane\altaffilmark{2} and Lev Titarchuk\altaffilmark{3,4}}

\altaffiltext{1}{Department of Physics, University of
Padova, via Marzolo 8, 35131 Padova, Italy; turolla@pd.infn.it}
\altaffiltext{2}{Mullard Space Science Laboratory, University College London,
Holmbury St. Mary, Dorking, Surrey, RH5 6NT, UK; sz@mssl.ucl.ac.uk}
\altaffiltext{3}{George Mason University/Center for Earth
Observing and Space Research, Fairfax, VA 22030 and US Naval Research
Laboratory, Code 7620, Washington, DC 20375-5352, USA; lev@xip.nrl.navy.mil }
\altaffiltext{4}{NASA/Goddard Space Flight Center, Greenbelt
MD 20771, USA; lev@lheapop.gsfc.nasa.gov}

\begin{abstract}

The effects of bulk motion comptonization on
the spectral formation in a converging flow
onto a black hole are investigated. The problem is
tackled by means of both a fully relativistic, angle-dependent
transfer code and a semi-analytical, diffusion-approximation method.
We find that a power-law high-energy tail is a ubiquitous feature in
converging flows and that the two approaches produce consistent results at
large enough accretion rates, when photon diffusion holds.
Our semi-analytical approach is based on an expansion in eigenfunctions of
the diffusion equation. Contrary to previous investigations based on the
same method we find that, although the power-law tail at really large
energies is always dominated by the flatter spectral mode, the slope of
the hard X-ray portion of the spectrum is dictated by the second mode
and it approaches $\Gamma=3$ at large accretion rate, irrespective of the
model parameters.
The photon index in the tail is found to be largely independent on the   
spatial distribution of soft seed photons when the accretion rate is
either quite low ($\la 5$ in Eddington units) or sufficiently high ($\ga
10$). On the other hand, the spatial distribution of source photons
controls the photon index at intermediate accretion rates, when $\Gamma$
switches from the first to the second mode.
Our analysis confirms that a hard tail with photon index
$\Gamma <3$ is produced by the up-scattering of primary photons onto
infalling electrons if the central object is a black hole.
\end{abstract}


\section{Introduction}
\label{intro}
The idea that photons may change their energy in repeated scatterings with
cold electrons in a moving fluid has been suggested more than 20 years ago
by \citet{pb:1981} and \citet{cl:1982}. This
process, often refereed to as dynamical (or bulk) Comptonization, is
completely equivalent to Comptonization by hot electrons
once the thermal velocity is replaced by the bulk velocity ${\bf v}$.
However, as already
noted by Cowsik \& Lee and Payne \& Blandford, it has to be
$\nabla\cdot{\bf v}\neq
0$, like in a converging flow, for the mechanism to produce a full-fledged
effect. If a photon interacts with electrons moving at uniform
speed its
energy is boosted by a factor $\sim \gamma^2=(1-v^2)^{-1}$, independently of
the number of scatterings. On the other hand, in a flow where
$\nabla\cdot{\bf v}\neq 0$ a photon typically
scatters on electrons with different velocity and the change of the local
rest frame introduces a differential effect. As shown by Payne \& Blandford
(see also \citealt{ntz:1993}),
if monochromatic radiation with $\nu=\nu_0$ is injected
at large Thomson depth in a spherical accretion flow the emergent spectrum
is broad, shifted to $\nu > \nu_0$ and a typical power-law tail appears at
high energies. For a power-law velocity law, $v\sim r^{-\beta}$, the
photon spectral index is correlated to the velocity gradient and becomes $3$
in free-fall. Dynamical Comptonization in
the non-relativistic limit was also investigated by
\citet{sb:1989} and \citet{mk:1992}, who considered a
spherical flow onto a neutron star and found that the choice of the inner
boundary condition affects the emerging power-law index substantially.
The competing action of dynamical and thermal Comptonization in a
semi-infinite medium at non-zero temperature was discussed by
\citet{col:1988} using the non-relativistic diffusion  approximation.
\citet{tmk:1996} extended Mastichiadis \& Kylafis' results
including the effects of electron recoil and thermal motion.
They demonstrated that the spectral power-law index goes to zero when
the Thomson depth in the flow becomes very large.

The potential importance of dynamical Comptonization in connection
with high-energy emission from compact X-ray sources has been
recognized since the very beginning. However, it was not until
the mid-90's that renewed interest in this topic was aroused by
two facts: the observational evidence that X-ray spectra from
Galactic black hole candidates (BHCs) may exhibit in the soft
(high) state a power-law tail which extends up to hundreds of
keV, and the introduction of the two-phase paradigm for accretion
flows onto black holes (\citealt{ct:1995}). In addition to
standard, geometrically-thin accretion disks (SSDs,
\citealt{ss:1973}), \citet{ct:1995} presented arguments for the
existence of a sub-Keplerian flow outside the disk. The popular
advection-dominated accretion models (ADAFs), initially
introduced for optically thin accretion flows onto black holes
(see e.g. \citealt{nmq:1999} for a review), are a particular
class of sub-Keplerian flows, namely ADAFs are also nearly
spherical and transonic close to the hole. In current models for
BHCs, a sub-Keplerian component is thought to exist along with
the SSD in the inner region of the accretion flow ($\la 100$
gravitational radii) and it may provide the conditions for making
dynamical Comptonization effective. \citet{ct:1995},
\citet{tmk:1996}, and \citet{etc:1996} were the first to point
out that in a realistic accretion flow the finite Thomson depth
at the horizon, $\tau_{s,H}$ produces a power-law spectral index
(which depends on $\tau_{s,H}$) flatter than 3, in general
agreement with the observed values (at least for $\dot m \la 5$,
as we will discuss later).  \citet{tzzn:1996}, solving
analytically the general-relativistic moment equations, confirmed
this result and provided a simple expression for the power-law
index as a function of $\tau_{s,H}$ in a free-falling medium.
\citet{tmk:1997}, hereafter TMK97, considered a non-relativistic
diffusion equation for the photon occupation number, including
dynamical and thermal effects. They derived both numerical and
(approximate) analytical solutions, and concluded that, while the
spectral index depends on the location of the inner boundary and
on the boundary conditions, it is not much sensitive to the
spatial and energy distribution of the primary photons. Several
numerical calculations, based on different approaches to the
solution of the transfer problem, have been presented so far
(relativistic moment method: \citealt{ntz:1993},
\citealt{ztt:1993}; fully relativistic transfer equation:
\citealt{ztne:1996}, \citealt{tz:1998};  Monte Carlo simulations:
\citealt{lt:1999}). \citet{st:1998}, following TMK97, were able
to reproduce the observed X-ray spectrum of several BHCs, assuming
that the accretion flow consists of two phases: a SSD and a radial
component which replaces the disk close to the hole. Thermal
photons emitted by the disk produce the observed soft emission at
a few keV and are in part up-scattered by the inflowing electrons
forming the power-law tail.

Numerical investigations based on Monte Carlo methods or
$\Lambda$-iteration schemes typically fail in reproducing cases
with relatively high $\dot m$, even well below the range where
diffusion approximation holds. Analytical investigations, on the
other hand, are supposed to reproduce the diffusion approximation
limit, but suffer from a series of limitations imposed by the
simplifying assumptions which need to be introduced. All of them
are based on the solution of the first two moment equations
neglecting, to a various extent, terms of higher order in $v$ and
higher order moments. This approach is justified at large depth,
but becomes questionable for $\tau_{s,H}\la 1$. Moreover, analytical
spectra have been often calculated for a monochromatic injection
of photons at a given radius, so they fail to answer to the
fundamental question of how the emergent spectrum depends on the
primary photon distribution, both in energy and in space. An
exception is the method presented by TMK97, that will
be used in this investigation.

Quite surprisingly, while
all studies seem to agree that a power-law tail forms at high
energies in a converging flow, the derived values of the spectral index
cover a considerable range. This can reflect either an inaccuracy
of the methods or an intrinsic dependence of the spectral index on some
(often implicit, even hidden) assumptions at the basis of the calculation,
or both. In the attempt to clarify this point \citet{pp:2000}
undertook a project aimed to a systematic exploration of the parameter
space by means of the numerical integration of the relativistic transfer
equation for a free-falling flow in a Schwarzschild spacetime. The results
reported in their first paper are in agreement with the single model of Zane
et al. (1996) but not with the analytical prediction of Turolla et al.
(1996).
The derived values of the spectral index are  similar to those
of Titarchuk \& Zannias and Laurent \& Titarchuk, but care must
be taken in comparing this calculation to previous ones because
\citet{pp:2000} use a different definition of $\tau_{s,H}$.

In this paper we clarify the role of the spatial distribution of
the source of input photons. This has been often overlooked or
misunderstood in the literature. In particular, we consider the
difference between diffuse and spatially concentrated sources. We
support our results with both a systematic numerical analysis and
a semi-analytical calculation. The first is carried out by solving
the transfer problem for a radially inflowing medium in a
Schwarzschild spacetime using the code described in
\citet{ztne:1996}. The latter is based on the method introduced
by TMK97. Numerical models have been computed up to large enough
values of $\tau_{s,H}$ to allow for a direct comparison with
analytical, diffusion-approximation results. We show that the
spatial distribution of seed photons is unimportant in fixing the
power-law index both al small and large accretion rates ($\dot
m\la 5$ and  $\dot m\ga 15$ for a scattering-dominated accretion
flow onto a black hole), while it influences the index at
intermediate accretion rates. However, unless the source is
extremely concentrated close to the horizon, the slope of
high-energy tail does not depend much on the primary photon
distribution. We also convincingly show that the photon index
approaches $3$ at large accretion rates, irrespective of the
input parameters.

The plan of the paper is as follows. In \S\ref{numres} we introduce
the numerical method for the solution of the relativistic kinetic equation
and present the results of numerical calculations. Approximate analytical
solutions in diffusion approximation are derived in \S\ref{analytic}
and compared with both our numerical models and results from previous
investigations. Discussion and  conclusions follow in \S\ref{discuss}.

\section{Numerical method and results}
\label{numres}

A fairly general technique for the numerical solution of the
relativistic transfer equation in spherical symmetry has been
presented and discussed in \citet{ztne:1996}. The method makes
use of the characteristics to reduce the comoving-frame transfer
equation to an ordinary differential equation for the photon
occupation number $f$ along the photon trajectories. To avoid any
confusion, we stress that $f=f(r,\mu,E)$, the cosine of the angle
between the photon and the radial directions $\mu$ and the photon
energy $E$ are all measured by the comoving observer (LRF), while
$r$ is the coordinate radius.

In the following we deal with a
spherical flow in a Schwarzschild spacetime and use units in
which $c=G=h=1$. The radial coordinate is in units of the
Schwarzschild radius, $r_S = 2M$, where $M$ is the mass of the
central source. For a Schwarzschild geometry Zane et al. derived
simple analytical expressions for $\mu = \mu(r,b)$ and
$E=E_\infty\epsilon(r,b)$, where $b$ is the ray impact parameter
and $E_\infty$ the photon energy measured by an observer at rest
at radial infinity. The transfer equation
\begin{equation}\label{transf}
\frac{df}{dr} = \frac{r_S {\cal G}(r,\mu,E,f)}{yE(\mu+v)}\,,
\end{equation}
where $y=\gamma\sqrt{1-r_S/r}$ and $\cal G$ is the source term, is then solved
for different values of the two parameters $b$ and $E_\infty$ to obtain
the specific intensity $I= 2 f E^3$. Ordinary $\Lambda$-iteration
is used to
reach convergence in case the source term contains scattering integrals.
The radiation moments (mean intensity $J_\nu$, flux $H_\nu$ and pressure
$K_\nu$, here $\nu$ is the photon frequency) are
evaluated by direct numerical quadrature over angles of the specific
intensity times the required power of $\mu$, at constant $E$ and $r$.
We assume conservative (i.e. Thomson) scattering in the
electron rest frame. Since we want to assess the effects
of dynamical Comptonization, we ignore thermal motion and take the
electron rest frame to coincide with the LRF (TMK97).

We assume free-fall so that $yv = r^{-1/2}$ which implicitly
gives  $v$ as a function of $r$. Denoting with $\dot M$ the
accretion rate, it follows from the rest mass conservation $\dot
M=4\pi r_S^2 c r^2\rho yv$, that the gas density scales as $\rho
=\rho_H r^{-3/2}$. Since $y\simeq 1$ in free-fall (hence $v\simeq
r^{-1/2}$), and introducing the accretion rate in units of the
Eddington rate $\dot m = \dot M/\dot M_E$ ($\dot M_E=L_E/c^2$),
the density at the horizon is related to $\dot m$ by $\dot m =
2\kappa_s\rho_H r_S$, where $\kappa_s$ is the scattering opacity.
The expression for the scattering depth in the flow follows
immediately and is $\tau_s = \int_r^\infty\kappa_s\rho r_S\, dr=
2\kappa_s\rho_H r_S r^{-1/2}=\dot m r^{-1/2}$. The effects of
bulk motion Comptonization on the emerging spectrum are governed
by the product of the scattering depth times the flow velocity
which gives the fractional energy change suffered by a photon
undergoing repeated scatterings before escaping (see e.g.
\citealt{ntz:1993}). In the present case $\tau_sv = \dot m
r^{-1}$ and the trapping radius, defined as the locus where
$3\tau_sv=1$, is located at $r_{trap} = 3\dot m$

We have computed several sequences of models for $\dot m$ in the range
$1\leq\dot m\leq 12$. All models include both electron scattering and
true emission/absorption.
In order to investigate the effects of the spatial
distribution of the input photons under a minimal set of assumptions, we
adopted an artificial opacity coefficient $\kappa_{a,\nu}$ defined as to
produce an absorption depth

\begin{equation}
\label{tauabs}
\tau_a  =\left\{\begin{array}{ll}
            \tau_{a,H} & \mbox{$r \le r_a$}\\
                & \cr
               \displaystyle{\tau_{a,H}\left(\frac{r_a}{r}\right)^n}
               & \mbox{$r > r_a$}
           \end{array}
          \right.
\end{equation}
where $\tau_{a,H}$ is the absorption depth at the horizon and $r_a$ and $n$
are adjustable parameters. With the above definition absorption is
color-blind, i.e. $\kappa_{a,\nu}=\kappa_a$. However, since we retain
Kirchhoff's law, the emission coefficient depends on frequency and is given by
\begin{equation}\label{emiss}
j_\nu = \tau_a r^{-1}B_\nu(T)
\end{equation}
where $B_\nu(T)$ is the Planck function at temperature $T$. We
then want to explore the effects of bulk motion Comptonization on
the emerging spectrum when $\dot m$ is varied, along various
sequences of models characterized by different spatial
distributions of the emissivity, eq. (\ref{emiss}). Particular
care must be devoted to the fact that the sequences should be
self-similar as far as the role of emission/absorption is
concerned. If this is not ensured the spectral behavior may be
influenced by the different interplay between scattering and
absorption when the accretion rate changes. The self-similarity
can be imposed in different ways: here we ask that the scattering
and absorption depths become equal at a given radius $r_c$
(hereafter the ``crossing radius''), which is the same for all
models. Using the expression for the absorption depth given by
(\ref{tauabs}) with fixed $n$, the condition $\tau_s(r_c) =
\tau_a(r_c)$ allows to derive $\tau_{a,H}$ for each value of
$\dot m$, $\tau_{a,H}=\dot m (r_c/r_a)^{n-1/2}r_a^{-1/2}$. In our
calculations we used $r_a=2.5$, $r_c=1.8r_a$ and $n$ in the range
$3\leq n\leq 7$. We note that both $r_a$ and $r_c$ should be of
order unity if the flow has to be scattering dominated up to small
radii. Provided this condition is fulfilled, we have checked that
varying them does not change the models much.

With this choice of the parameters the absorption depth at $r_c$
is always larger than unity for $\dot m \geq 1$, so all models
have an inner core which is optically thick to true absorption.
Beyond $r_c$ both the source of seed photons and the true opacity
rapidly decay with radius, the distributions becoming sharper with
increasing $n$, thus in this region bulk motion Comptonization is
the most efficient process. Figure \ref{radii} illustrates the
dependence of the relevant length-scales on the accretion rate and
Figure \ref{emi} shows the radial variation of the
frequency-integrated emission coefficient for different values of
$n$ and $\dot m$. Because of our assumption of self-similarity,
upon normalization all curves with the same $\dot m$ coincide in Fig.
\ref{emi}. All models have been computed on a radial grid which
covers the range $0.2\leq \log r\leq 6$ and is equally spaced in
$\log r$. Although the actual integration of eq. (\ref{transf})
along the rays has been carried out on a much finer grid, the
radiation intensity was stored for 40 values of the radial
coordinate. Since our main goal is to investigate the effects of
dynamical Comptonization, we consider here an uniform temperature
medium. The exact value of $T$ in unimportant and only fixes the
scales of both the photon energy and the intensity (the latter
because we have thermal emission). We used 30 energy points,
taken to coincide with Gauss-Lobatto quadrature abscissae, in the
range $0.3\leq E/kT\leq 20$, plus additional ten points both
below and above the two limits (see \citealt{ztne:1996} for
details). In our scheme the angular resolution is fixed by the
number of rays along which the transfer equation is integrated
and is not constant along the radial grid. With our present
choice of the parameters the minimum number of $\mu$ points
(which happens at the photon radius or at the inner boundary if
the latter is larger) is 20. Because $\tau_a>\tau_s>1$ for
$r<r_a=2.5$ and the absorption opacity in independent on
frequency, the spectrum below $r_a$ is blackbody (see eq.
[{\ref{emiss}]). For this reason we decided to put the inner
boundary just outside the photon radius, at $r_{b}\simeq 1.6$.
This speeds up the calculation since the transfer equation needs
not to be solved along the trapped photon trajectories. Standard
boundary conditions for a non-illuminated medium have been used:
$f(r_{b}, \mu>0) = B_\nu(T)/E^3$ and $f(r_{out}, \mu<0) = 0$,
where $r_{out}$ is the outer boundary of the integration domain.

The emerging spectrum for different $\dot m$ is shown in Figure \ref{spectra}
for the three cases $n=3,\, 4,\, 5$. The appearance  of a high-energy
power-law
tail is clearly seen comparing the emergent photon distribution with the
blackbody spectrum (the dashed line in Fig. {\ref{spectra}). The spectral
index depends both on $\dot m$ and on the particular sequence of models.
The derived values of the photon index $p$ are plotted as a function of
$\dot m$ for different $n$ in
Figure \ref{index}.

\section{Approximated Analytical Solutions}
\label{analytic}

The numerical results presented in \S \ref{numres} made
evident a dependence of the spectral index on the properties of
each particular sequence, at least in the range of $\dot m$ that
has been spanned. In order to address this point further, we
investigate systematically the properties of the analytical
solutions presented by TMK97. We then present an application of the method
to cases relevant to the models presented in \S \ref{numres}. Although
this
approach strictly holds only for an isotropic radiation field, we
will show that it correctly describes the numerical sequences for
$\dot m$ not too close to unity and helps understanding the
dependence of the numerical models on the various input
parameters.

\subsection{The analytical solution}
\label{ansol}

By following TMK97, we write the kinetic equation for the
angle-averaged photon occupation number $n=n (r, \nu)$ [their eq.
(14)] as
\begin{equation}
\label{diffeq}
\tau \derss {n} {\tau} - \left (\tau + {3 \over 2} \right
)  \der { n} {\tau} - {1 \over 2} x  \der {n} {x} = -{ \dot m \over 2} {
j_\nu \over \rho \kappa_s}\, ,
\end{equation}
where $\tau = (3/2) \dot m/r$, $x = h \nu /kT$. As in TMK97,
second order terms in $v$ have been neglected and we restricted
to the case in which the source function is the product of a
purely spatial part, $S(\tau)$, and a purely energy-dependent
part, $g(x)$. Moreover, we did not include thermal Comptonization
and treated the scattering as elastic in the electron rest frame.

The eigenfunctions or the space (radial) operator which are well behaved
($\sim r^{-2}$
for  $r\to \infty$) are given by
\begin{equation}
R_k = C \tau^{5/2} \Phi \left ( -\lambda_k^2 + 5/2, 7/2, \tau \right )
\end{equation}
where $\Phi (a,b,z)$ is the confluent hypergeometric function
(Kummer's function, see \citealt{as:1970}), $C$ is a constant and
the eigenvalues $\lambda_k^2$ are the roots of the equation 

\begin{equation}
\label{roots}
p \Phi\left (-\lambda_k^2 + {5 \over 2} , {9 \over 2},\tau_b \right )
+ q \Phi \left (-\lambda_k^2 +  {3 \over 2},  {9 \over 2}, \tau_b \right )=0
\end{equation}
with
\begin{eqnarray}
p& = &
\left [ {5 \over 2} - \left ( 2 \lambda_1^2 +\epsilon
\right )
{\tau_b \over 3} \right ] \left (- \lambda_k^2 +{3 \over 2}+\tau_b \right)+
\nonumber\cr
& &  \tau_b \left (-2 \lambda_k^2+ {1 \over 2} + \tau_b \right ) \, ,
\nonumber \cr q & = & \left [{5 \over 2} - \left (2 \lambda_1^2
+\epsilon -3 \right){ \tau_b \over 3} \right ] \left (
\lambda_k^2 +2 \right ) \, ; \nonumber
\end{eqnarray}
here $\epsilon = -3 (1 -A) \sqrt{r_b}/[2 (1 + A)]$, $\tau_b$ and $r_b$ are
depth and the
radius of the inner boundary, respectively, and $A$ is the albedo at $r_b$. We
note that eq. (\ref{roots}) is just a rearrangement of eq. (A6)
of TMK97.

Expanding the spatial part of the source function over the
complete set of eigenfunctions $R_k$
\begin{equation}
S(\tau)  = \sum_{k=1}^\infty c_k R_k(\tau)
\, , 
\end{equation}   
and looking only for separable solutions of the form

\begin{equation}
n(\tau, x) = R(\tau)N(x)=\sum_{k=1}^\infty a_k R_k(\tau) N_k(x)\,,
\end{equation}
gives

\begin{equation}
N_k(x) =  2{c_k \over a_k}
x^{-2 \lambda_k^2} \int_0^x t^{2 \lambda_k^2 -1}g(t) dt
\equiv {c_k \over a_k} \hat N_k(x)\,.
\label{nk}
\end{equation}
The emergent luminosity is then expressed as 

\begin{equation}
\label{emlum}
L(\tau=0, x) \propto  x^3 \sum_{k=1}^\infty c_k \hat N_k(x)\,.
\end{equation}
We note that the coefficients $a_k$ do not enter expression (\ref{emlum}).
The emergent luminosity only depends on the $c_k$'s, i.e.
on the spatial distribution of the source function. These
coefficients may be readily evaluated as integrals of $S(\tau)$ times a 
weighting function (see TMK97, Appendix B).
The source function corresponding to the numerical models
presented in \S \ref{numres} is $S(\tau) \propto \tau^{n-1/2}$
for $r \geq r_a$ where scattering is dominant.

In order to investigate the behavior of the spectral index, we
evaluated $L(\tau=0,x)$ in the range $0.1 \leq x \leq 500$, for
different values of $\tau_b$ and $n=3,\, 4,\, 5$, assuming that
$g(x)\propto [\exp(x)-1]^{-1}$ (i.e. the primary photon spectrum is
blackbody). We use here the same
notation as in TMK97, where the eigenvalues were labeled according
to their asymptotic value at $\dot m \gg 1$. In particular, the
first eigenvalue $\lambda_1^2$ is the smallest root of eq.
(\ref{roots}). Since the different roots never cross each other
when $\dot m$ decreases, the hierarchy of the eigenvalues can be
derived from their asymptotic values at large $\tau_b$. As shown
by TMK97, in this limit the roots are
\begin{eqnarray}
s_1 & \sim & {3 \over 2} + {3 \over 4} \left ( { 1 - A \over 1
+ A} \right ) \sqrt{r_b}\,,\nonumber\cr
s_k & \sim & { 2 k + 1 \over 2}, \qquad k\geq 2\nonumber
\end{eqnarray}
out of which only the smallest one is meaningful and should be then inserted
back into eq. (\ref{roots}) to compute the $\lambda_k^2$'s with $k\geq 2$.
As the previous expressions shows, it is $\lambda_1^2 = s_1$ if
$3/2 + (3/4) (1 - A)(1+ A)^{-1} \sqrt{r_b} < 5/2$ while
$\lambda_1^2 = s_2$ otherwise. Once eq. (\ref{roots}) is solved with the 
appropriate value
for $\lambda_1^2$, the two sequences of the eigenvalues at large
$\tau_b$ turn out to be
\begin{eqnarray}
\lambda_1^2  & \sim & {3 \over 2} + {3 \over 4} \left ( { 1 - A \over 1
+ A} \right ) \sqrt{r_b}\,,\nonumber\cr
\lambda^2_k & \sim & { 2 k + 1 \over 2}, \qquad k\geq 2\nonumber
\end{eqnarray}
for $3/2 + 3/4 (1 - A)/(1+ A) \sqrt{r_b} < 5/2$ and
\begin{equation}
\lambda_k^2 \sim {2k+3\over 2}\,\qquad k\geq 1\nonumber
\end{equation}
for $3/2 + (3/4) (1 - A)(1+ A)^{-1} \sqrt{r_b} > 5/2$.
Since $\hat N_k(x)\sim x^{-2 \lambda_k^2}$ for $x \gg 1$, the
flatter spectral mode corresponds to the smaller $\lambda_k^2$.

It is natural to expect that terms of
higher order do not contribute significantly to the series
(\ref{emlum}) at large frequencies. Naively, one might be tempted
to conclude that the
value of the spectral index in the power law tail is only
dictated by the smallest eigenvalue (the same conclusion
is in fact at the basis of most previous investigation, see e.g.
TMK97). However, while terms with $k>2$ never significantly
contribute to the high-energy tail, the dominant mode can be
either the first or the second, depending on the
parameters of the model. For $\lambda_1^2\to 5/2$ the first mode
is always the dominant one, but when 
$\lambda_1^2\to 3/2 + (3/4)
(1 - A)(1+ A)^{-1} \sqrt{r_b} < 5/2$, 
it dominates
only in a limited range of relatively small $\dot m$. This is illustrated
in Figure \ref{kk0}a, b, which shows the first nine 
terms of the series (\ref{emlum}) for $A=0$, $r_b=1$, $n=5$ and two 
different values of $\dot m = 2\tau_b r_b/3$. At
relatively low $\dot m$
(Fig.~\ref{kk0}a) the mode $k=1$ is indeed the dominant one, but
when $\dot m$ increases (Fig.~\ref{kk0}b) it is the $k=2$ term
which gives the larger contribution over many decades in
frequency. The main reason is that the expansion coefficient
$c_1$ goes to zero exponentially fast when $\tau_b$ increases.
This has been already noted by TMK97
(see also \citealt{mk:1992}) in discussing the equivalence
of their solution to that of \citet{pb:1981} for $\tau_b
\rightarrow \infty$. However they did not point out that, as a
consequence, the eigenvalue $\lambda_1^2$ does not represent the
spectral index at large $\dot m$. The main consequence is the
appearance of a range in $\dot m$ over which the spectral index
makes a ``transition'' from the first to the second mode.
This is illustrated in Figure.~\ref{trans} which makes evident 
that the spectral index in the transition region do depend on the spatial
distribution of the source. We stress that all these considerations
do not reflect the behavior of the power-law tail in the limit
$x\to\infty$. At really large energies, the spectral index is always
fixed by $\lambda_1^2$, as it can be guessed from Figure~\ref{kk0}b.
However, the observationally accessible
part of the high-energy tail is indeed dictated by $\lambda_2^2$.

Figure~\ref{rb}a illustrates the dependence of the spectral index on
$r_b$. As it can be seen, the variation with $r_b$ is not
monotonic. At low $\dot m$ spectra tend to be harder as the inner
boundary moves closer to the horizon, while at larger $\dot m$
the behavior of the spectral index is more complicated. Models
with different $r_b$ approach $2\lambda_2^2-2$ in correspondence
of different values of $\dot m$, and beyond $\dot m \approx 10$
spectra might be softer at smaller $r_b$. It is also worth to
stress that, since $\dot m = 2\tau_b r_b/3$ and diffusion
approximation holds for $\tau_b>1$, the larger is $r_b$ the
larger is the value of $\dot m$ at which the analytical solution
is not trustworthy any more. The dependence on the albedo $A$ is
shown in Figure~\ref{rb}b. Again, all spectral indices tend toward
a common asymptotic value which is constant and is given by the
second mode. However, before this limit is reached, the dependence
on $A$ is monotonic. The larger is $A$ at the inner boundary, the
harder is the emergent spectrum, in agreement with previous findings
(see \citealt{mk:1992} and TMK97).

\subsection{Comparison with numerical models}

We are now in the position to reconsider the behavior of the
spectral index in the numerical sequences in the light of
the results discussed in the previous section. The
overall situation described by the numerical models is, of course,
different from that assumed in the analytical solution, in
particular because in the former a) the flow has an inner core,
optically thick to true emission/absorption,  b) true
absorption is consistently accounted for through Kirchhoff's law, and
c) relativistic effects are correctly included.

The first consequence is that the numerical sequences we have
computed are far from being evaluated at a given, fixed value of
both $r_b$ and $A$. However, a comparison between numerical and
analytical models can be attempted, assuming that $r_b$ coincides
with the boundary of the region in which scattering is the main
source of opacity, $r_{abs}$, where the absorption depth equals
unity [see \S \ref{numres} and eq. (\ref{tauabs})]. In the
region between $r_a$ and $r_{abs}$ the role of true absorption is
certainly non-negligible and affects the emerging spectrum
together with Comptonization. We stress that this is an
oversimplification, since it is not possible to define an inner
boundary in our numerical models in the same way as it was done
in \S \ref{ansol}. This implies also that the value of the
albedo at the inner boundary is ill-defined. On the other hand,
in all our models the region $r \la r_{abs}$ is optically thick
to absorption and is at constant temperature. Under these conditions, we
expect the ingoing flux at the boundary to be substantially larger
than the outgoing one. This amounts to say the inner part of the flow is
illuminated from above and the value of the albedo should
consequently be small, $A\la 0.1$.

In Figure ~\ref{comp} we show the
comparison between numerical and analytical sequences, the latter
computed taking $r_b = r_{abs} (\dot m)$, while $A$ is constant along
each sequence. For
$\dot m \ga 4$ the results derived in diffusion approximation are
in very good agreement with those computed with the full
angle-dependent code for reasonable values of the albedo. As
expected, the agreement becomes worse at low $\dot m$ where the
diffusion approximation starts to break down.

\section{Discussion and Conclusions}
\label{discuss}

Previous investigations on bulk motion Comptonization studied
the properties of the emerging spectrum, in particular as
far as the formation of a hard, power-law tail is concerned. In this
respect, two issues are of the highest relevance in connection with
observations of Galactic X-ray binaries (XRBs) in the high/soft state.
The first is whether the appearance
of a high energy power-law tail is a ubiquitous feature of converging
accretion flows, the second is to which extent the spectral index is
independent on the details of model, mainly on the spatial and energy
distribution of the seed photons.  TMK97, \citet{tz:1998} and \citet{lt:1999}
have shown that, for the cases they have examined,
a power-law tail is always present and the spectral index is almost
insensitive to source distribution. Our present results strengthen these
conclusions. At the same
time, we further clarify the properties of the spectral index and its
dependence on the detail of the source.

The main conclusion we derived in \S \ref{ansol} (see in particular
Fig. \ref{trans}) is that the photon index $\Gamma$ switches from
$\Gamma = 2\lambda_1^2-2$ to $\Gamma = 2\lambda_2^2-2$ as the accretion
rate increases. The location and width of the transition region depends in
general on the boundary conditions, i.e. on $r_b$ and $A$ (see e.g.
Fig. \ref{rb}b). For example, for the case $r_b=1$, $A=0$, the transition
occurs for $5\la\dot m\la 15$ while in solutions with larger $r_b$ the
transition region is wider and shifted toward larger $\dot m$.
The properties of the spectral index in the different ranges of
$\dot m$ (below, within and above the transition region) are dramatically
different, as it is shown by both numerical and analytical
results.

We now discuss this point in more detail by taking as
representative the case $r_b=1$, $A=0$, illustrated in Figure \ref{trans}.
As we can see, for low accretion rates, $\dot m \la 5$, the photon index
$\Gamma$ is indeed independent on the source spatial distribution and is
related only to the first eigenvalue, $\Gamma = 2\lambda_1^2 -2$. Although
$\lambda_1^2$ does
depend on both the albedo and the location of the inner boundary, its value
is always the same for a flow which is absorption thin all the way down to the
horizon.
This is likely to reproduce the situation in a black hole XRB, in
which most of the material is carried inwards by a standard disk and only
a small fraction accretes roughly spherically.
The different behavior of our numerical models is not in
contradiction with this statement. In fact, an important
point to realize is that, owing to the presence of an absorption thick
core, the range of $\dot m$ explored numerically
falls below (or at most at the beginning of) the transition
region.
The value of the photon index (which, in this range, is dominated
by $\lambda_1^2$) changes because of a change in both the
location of the inner boundary ($r_{abs}$ is a function of $\dot m$) and
the albedo.
Therefore, the variation of $\Gamma$ with $n$ (i.e. with the
photon source) seen in our numerical sequences is not
directly connected with the change of the source itself, but is largely
due to the change in the boundary conditions.

The situation is different in the transition region, i.e.
for intermediate values of the accretion rate ($5\la\dot m\la 15$ in the
case $r_b=1$ and $A=0$). Now the
spectral index genuinely depends on the spatial
distribution of primary photons. This
is true irrespective of the assumed boundary conditions, i.e. all
models are expected to exhibit a range in $\dot m$ (which can vary
depending on $r_b$ and $A$) over which the shape of $S(\tau)$ is important in
determining $\Gamma$ (see Fig. \ref{rb}b).
It should be noted, nevertheless, that $\Gamma$ depends rather
weakly on the spatial distribution, as it can be seen from Fig. \ref{trans}.
As a consequence, spatially concentrated photon sources (like the ones we
used here) or more diffused ones (as that employed by TMK97) give raise
to spectral indices which are not sensibly different. This implies that
$\Gamma$ is about the same for primary photons produced in the disk (diffused
source) or in the inner, denser region of an ADAF (concentrated source).

For large accretion rates (i.e. $\dot m\ga 15$ for
$r_b=1$ and $A=0$), the spectral index
quickly approaches $\Gamma=2\lambda_2^2-2$ and becomes again
independent on the source distribution. Since the value of $\dot m$ at
which the transition from $2\lambda_1^2-2$ to $2\lambda_2^2-2$
is completed is larger than the value at which $\lambda_2^2$ attains its
asymptotic value ($\dot m\sim 10$), the index is also
insensitive to both $r_b$ and $A$, and it is $\Gamma\sim 3$.
The same considerations apply to solutions with larger $r_b$ (see Fig.
\ref{rb}b).

We turn now to the comparison of  present results with those derived
in previous investigations for the spectral index and its dependence
of the accretion rate. Here we are concerned only with the case of
$A=0$ and $r_{b}=1$, that is to say we assume that the
accreting object is a black hole and true emission/absorption can be
neglected. The results are summarized
in Fig. \ref{compar}, where the spectral index is shown as a function of
the accretion
rate. As it can be seen, the major differences are at low values of
$\dot m$ where the various approaches give extremely different results.
We note first that methods based on the solution of the full (angle-dependent)
transfer problem provide quite similar (albeit not exactly equal) results
with $\Gamma$ a monotonically decreasing function of $\dot m$.
Relativistic diffusion approximation predicts much softer spectra and
$\Gamma$ is now monotonically increasing with the accretion rate.
Non-relativistic diffusion produces somehow an intermediate result in which
the spectral index exhibits a minimum.

An obvious point to recall is that
diffusion-limit solutions are valid only for
$\tau\gg 1 $, so they are bounded to fail when $\dot m\sim \tau_b\geq \tau$ is
a few or less. Moreover, the relativistic analysis by \citet{tzzn:1996}
strictly speaking applies only below the trapping radius. Clearly, for
small accretion rate the right answer is that
provided by the full transfer calculation. However, the comparison of the
various curves in Fig. \ref{compar} shows that the different behavior of
the index
at small $\dot m$ is not a direct effect of general relativity and hence
of the presence of the horizon. Would this be true one should expect
the relativistic diffusion approximation
to describe
much better the solution. Also, in free-fall gravity and dynamics cancel
each other at a given radius, so no relativistic effects are expected in the
flow local rest frame. This argument is not against the claim that the
detection of a power-law tail with index $\sim 2.5$--3 is indicative of
a black hole rather than a neutron star. The reason for this, nevertheless, 
are not
general relativistic effects but the fact
that only a black hole provides the conditions under which a converging
flow may reach $\tau_b\ga 1$. This is not possible for a neutron star,
because of the much higher accretion efficiency. The presence of a solid
crust at $\sim 3r_g$ implies that all the kinetic energy
of the inflowing material is released upon impact with an efficiency
$\approx 1/6$. This means that $\dot m\sim 1$ is enough to produce a
near-Eddington luminosity which slows down the flow until a settling solution
is established (see e.g. \citealt{mil:1990}; \citealt{ztt:1993}).
As a consequence, $\tau_b^{NS}\ll\tau_b^{BH}$ at the same
$\dot m$ because the velocity is much smaller. Increasing $\dot m$ is of no
avail since, as
\citet{ztt:1993} have shown, $\tau_b$ reaches $\sim 1$ for $\dot m\sim
3.5$ then starts to decrease. Consequently, for a spherically accreting
neutron star a power-law tail may indeed form but
it is always steeper (softer spectrum) than in black holes. It should be
realized also that the inner part of the accreting flow, close enough
to the star crust, is bound to be effectively thick, so assuming that
the surface acts as a reflecting boundary does not appear entirely realistic.

Finally, a comment of the asymptotic value of the index for $\tau_b\to\infty$
is in order. Both relativistic and non-relativistic diffusion calculations
give an asymptotic index $\Gamma=3$. The fact that the relativistic curve of 
\citet{tzzn:1996} approaches
3 at much larger values of $\dot m$ is related to the particular source
function they used. In fact, monochromatic photons were
assumed to be injected at a single radius very close to the horizon. 
The behavior of the index, as discussed previously, depends on the spatial
distribution of the source and the transition region is  wider and shifted
at larger $\dot m$ as the source is more and more concentrated. This shows
once more that
there are no substantial differences between relativistic and non-relativistic
diffusion, once proper allowance for the spatial
distribution of seed photons is made. 

In concluding, we would like to stress that
present results can not be directly compared with observations. As
several investigations have shown (e.g. \citealt{st:1998}; \citealt{st:1999};
\citealt{boroz:1999}), modeling the X-ray spectrum
of black hole binaries requires the inclusion of both thermal and
dynamical comptonization. Moreover, we neglected electron recoil 
(e.g. we assumed
elastic scattering in the electron rest frame),
so no high-energy cut-off is present in our models. If, as it seems
reasonable, the estimates derived by TMK97 and \citet{lt:1999} apply also
in the present case, a cut-off at energies $\approx 300-400$ keV is expected.
The lack of a break in high-energy spectra of black hole sources may
be suggestive of an alternative origin of the power-law tails, 
e.g. comptonization
by thermal/non-thermal electrons (e.g. \citealt{gier:1999}). Up to now,
however, observations
are not compelling in this respect (e.g. \citealt{zsky:2001} and
\citealt{st:2002} for OSSE observations of GRS 1915+105). Also,
\citet{lt:2001} found a high photon compactness near the horizon due to
relativistic ray bending. This should lead to pair production and ultimately
to an extension of the power-law tail to energies $\approx 1$
MeV. They have also shown that the power-law part of the spectrum at
energies less than $\approx$ 100-200 keV  is not affected by nonlinear
photon-electron interactions.

\section*{Acknowledgments}

Work partially supported by the Italian Ministry for Education, University,
and Research (MIUR) through grant COFIN-2000-MM02C71842. LT is grateful to
Kinwah Wu  for fruitful discussions during his visit to MSSL.

\clearpage

\begin{figure}
\includegraphics[width=6in,angle=0]{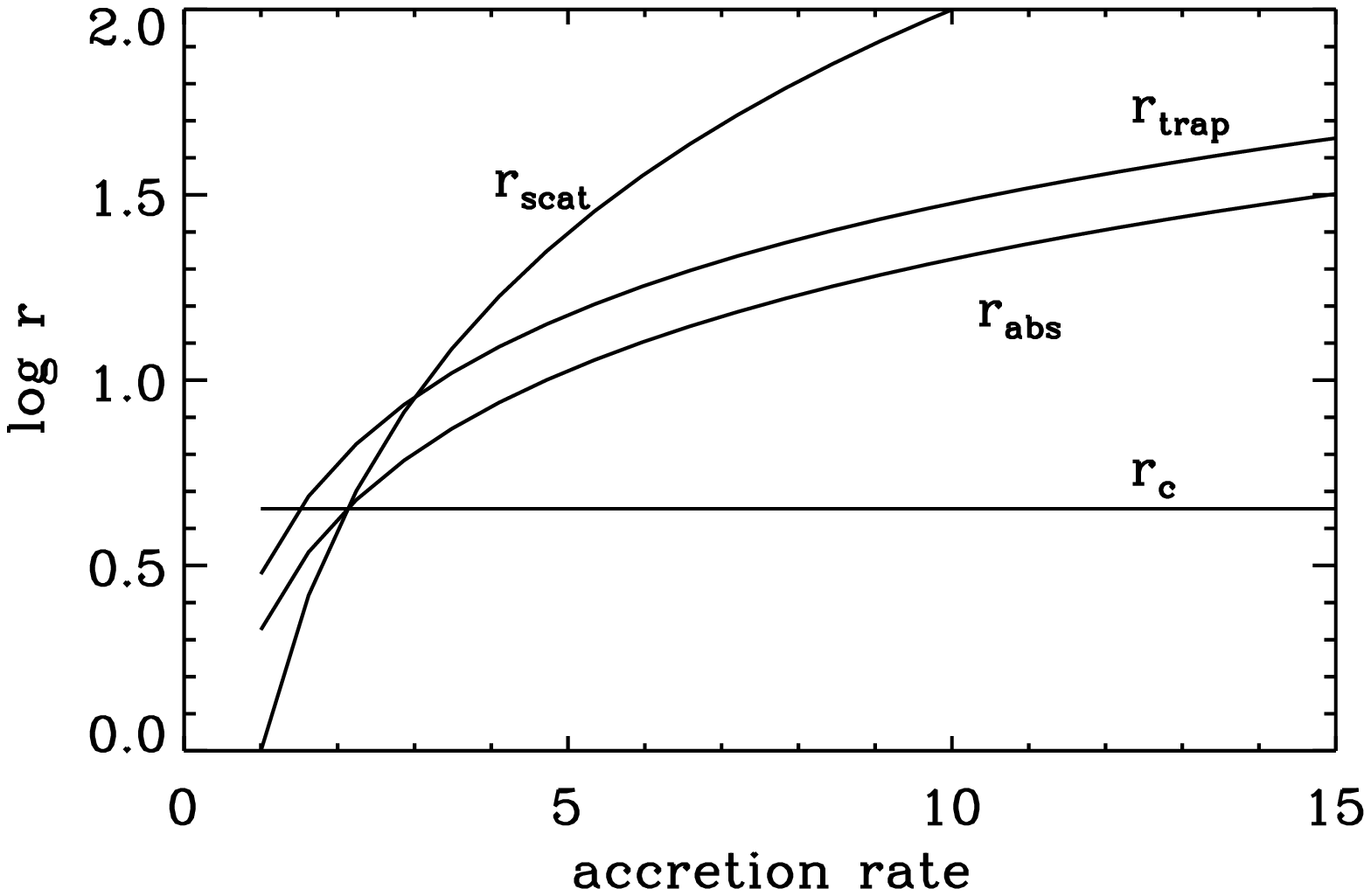}
\caption{\label{radii}The trapping radius $r_{trap}$, the crossing radius
$r_c$ and the scattering/absorption photospheric radii, $\tau_s(r_{scat})=
\tau_a(r_{abs})=1$, for different $\dot m$ and $n=4$.
}
\end{figure}

\begin{figure}
\includegraphics[width=6in,angle=0]{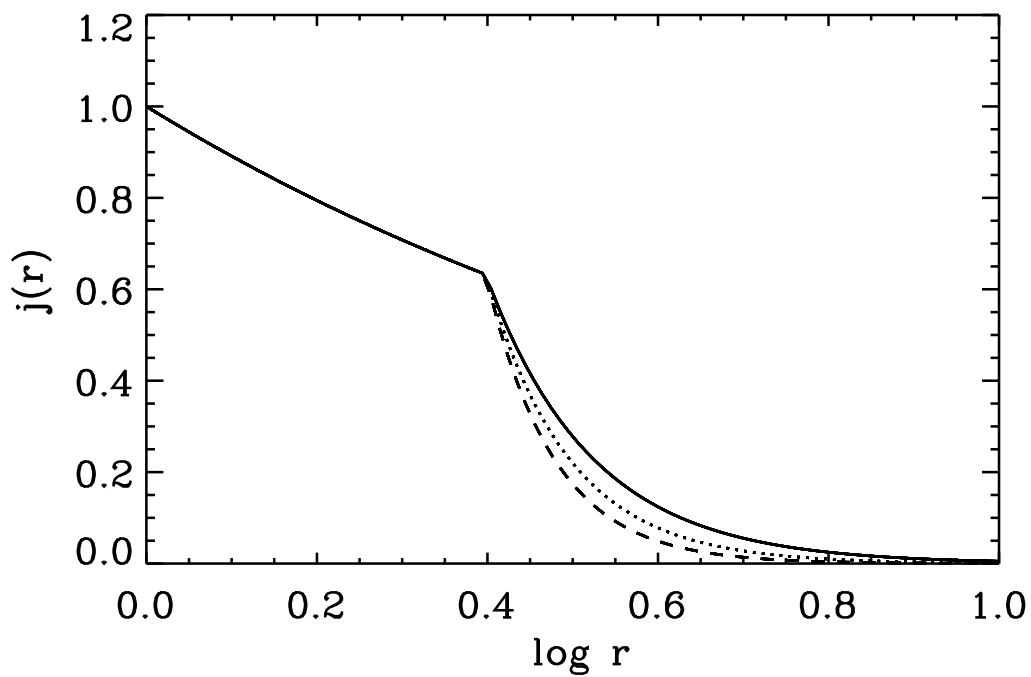}
\caption{\label{emi}The radial dependence of the frequency-integrated
emissivity for $n=3$ (full line), $n=4$ (dotted line) and $n=5$ (dashed
line); $j(r)$ has been normalized to its value at the inner radial point.
 }
\end{figure}

\begin{figure}
\includegraphics[width=6in,angle=0]{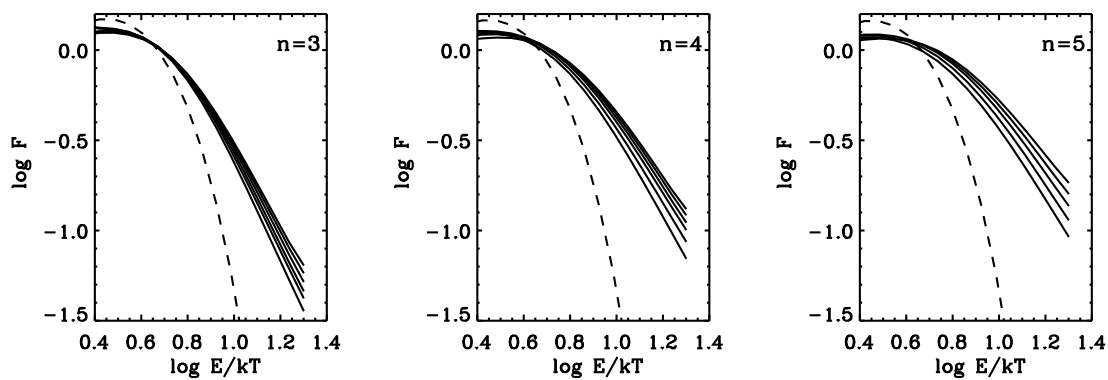}
\caption{\label{spectra}The emerging spectrum for $n=3,\,4,\,5$ and
different values of $\dot m$. In each panel $\dot m$ increases step 1
from bottom to top curve. Only the high-energy part of the spectra is
shown to visualize better the power-law tail. The dashed line is the
blackbody spectrum. Flux is in arbitrary
units and spectra have been normalized so that they coincide at the
lowest energy point (not shown).
 }
\end{figure}

\begin{figure}
\includegraphics[width=6in,angle=0]{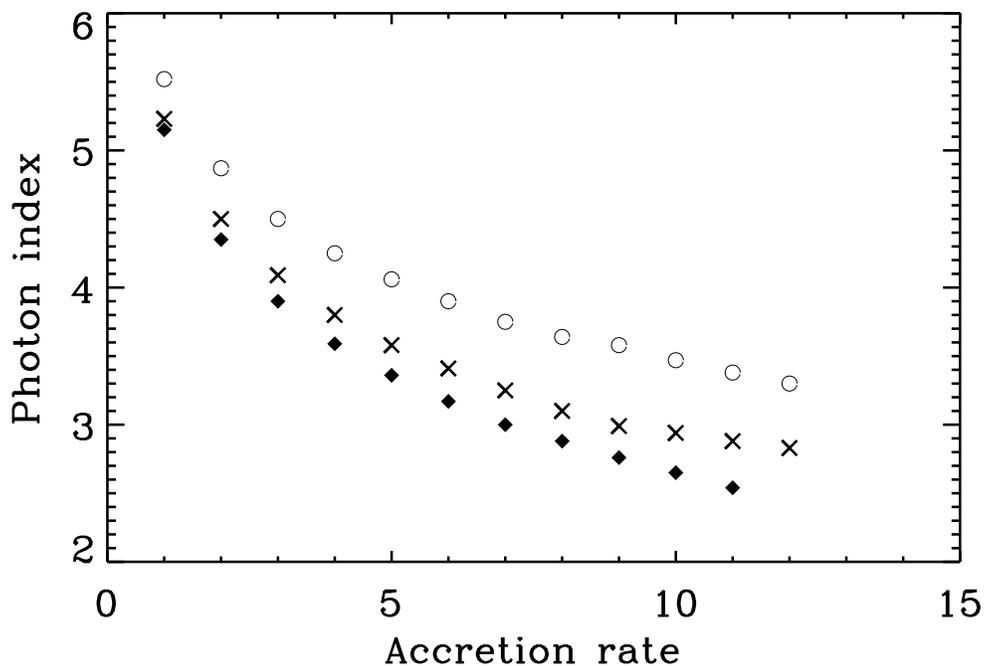}
\caption{\label{index}The photon index as a function of the accretion
rate for the different models discussed in the text: $n=3$
(circles), $n=4$ (crosses) and $n=5$ (diamonds).
 }
\end{figure}
\begin{figure}
\includegraphics[width=6in,angle=0]{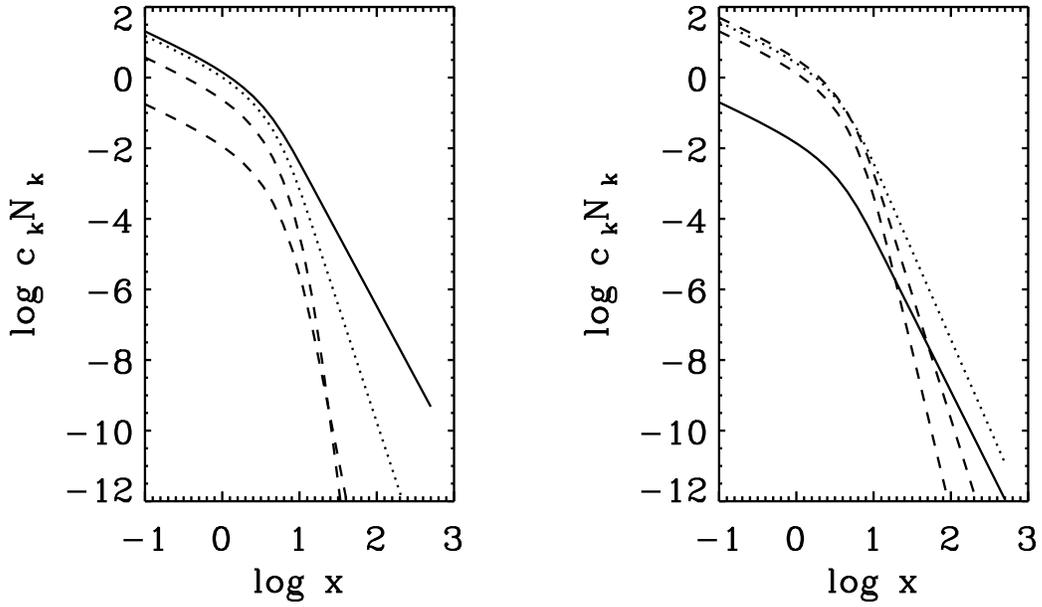}
\caption{\label{kk0}The first nine terms of the series for the
emergent flux vs. energy for $A=0$, $r_b=1$, $n=5$ and: a) $\tau_b
=6$ ($\dot m = 4$, left panel); b) $\tau_b
=20$ ($\dot m =13$, right panel). The full line corresponds to $k=1$, the
dotted line to $k=2$ and the dashed lines to $k=3,\, \ldots,\, 9$,
(from top to bottom). }
\end{figure}

\begin{figure}
\includegraphics[width=6in,angle=0]{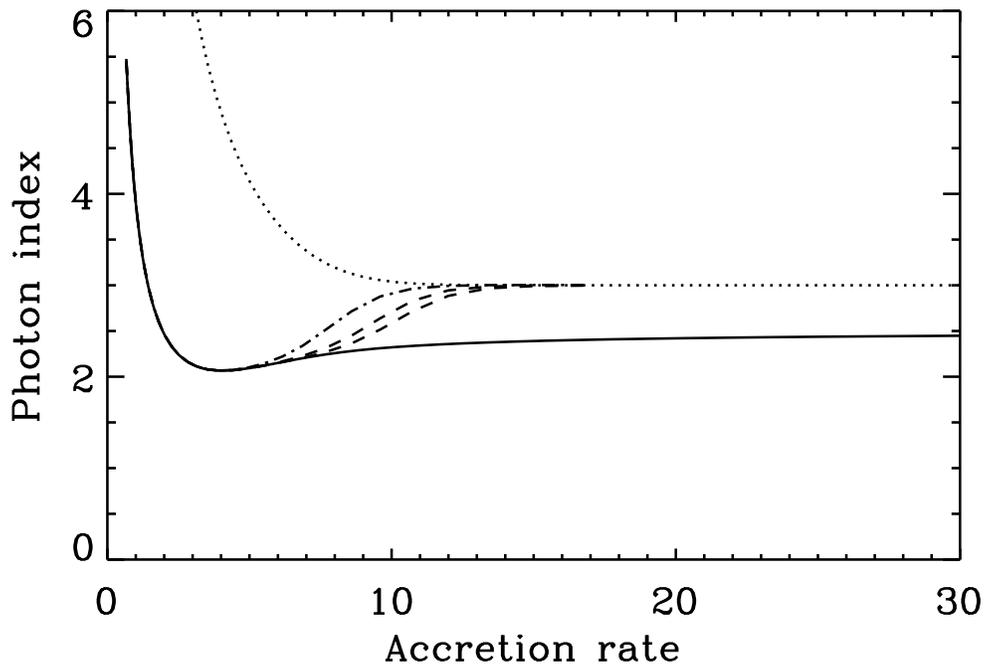}
\caption{\label{trans} The spectral index, derived from eq.
(\ref{emlum}), vs. $\dot m$: dashed lines are for $n=4$ (upper
curve) and $n=5$ (lower curve), the dash-dotted line corresponds to
TMK97 source distribution [their eq. (21)]. The full and dotted
lines represent $2\lambda_k^2-2$ for $k=1$ and $k=2$,
respectively. The presence of a transition region where the
spectral index switches from the first to the second eigenvalue is
evident; here $A=0$ and $r_b=1$.
 }
\end{figure}

\begin{figure}
\includegraphics[width=6in,angle=0]{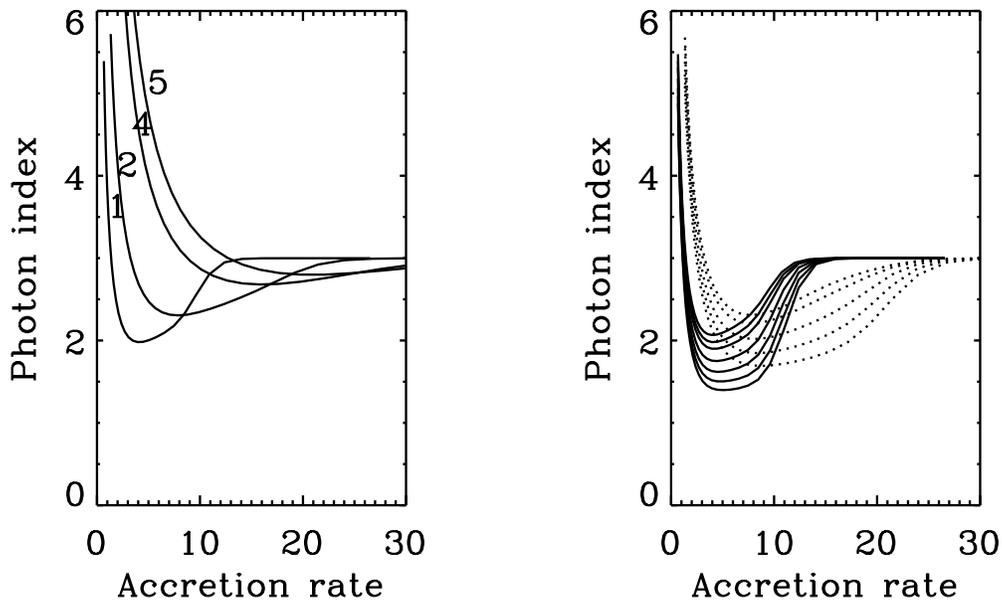}
\caption{\label{rb} The spectral index vs. $\dot m$ for $n=4$:
a) different values of $r_b$ (the labels on the curves) and $A=0.05$, 
(right panel); b) different values of $A$ and two values of $r_b$ (right
panel); here  $A=0, 0.05, 0.1, 0.2, 0.3, 0.4, 0.5$ ($r_b=1$,
full lines, from
top to bottom) and $A=0.05, 0.1, 0.2, 0.3, 0.4$ ($r_b=2$, dotted lines,
from top to bottom).
 }
\end{figure}

\begin{figure}
\includegraphics[width=6in,angle=0]{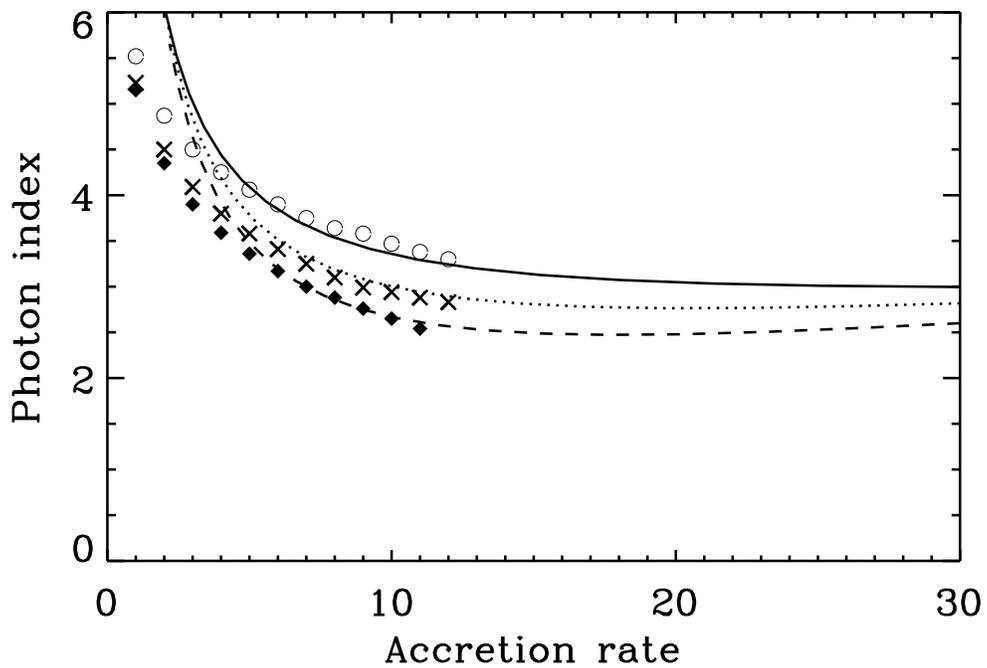}
\caption{\label{comp} The photon index for the numerical
sequences discussed in \S\ref{numres} compared with the analytical estimate
from eq.~(\ref{emlum}) with $r_b =r_{abs}$ for: $n=3$, $A=0.05$ (full
line), $n=4$, $A=0.1$ (dotted line), $n=5$, $A=0.2$ (dashed line).
 }
\end{figure}
\begin{figure}
\includegraphics[width=6in,angle=0]{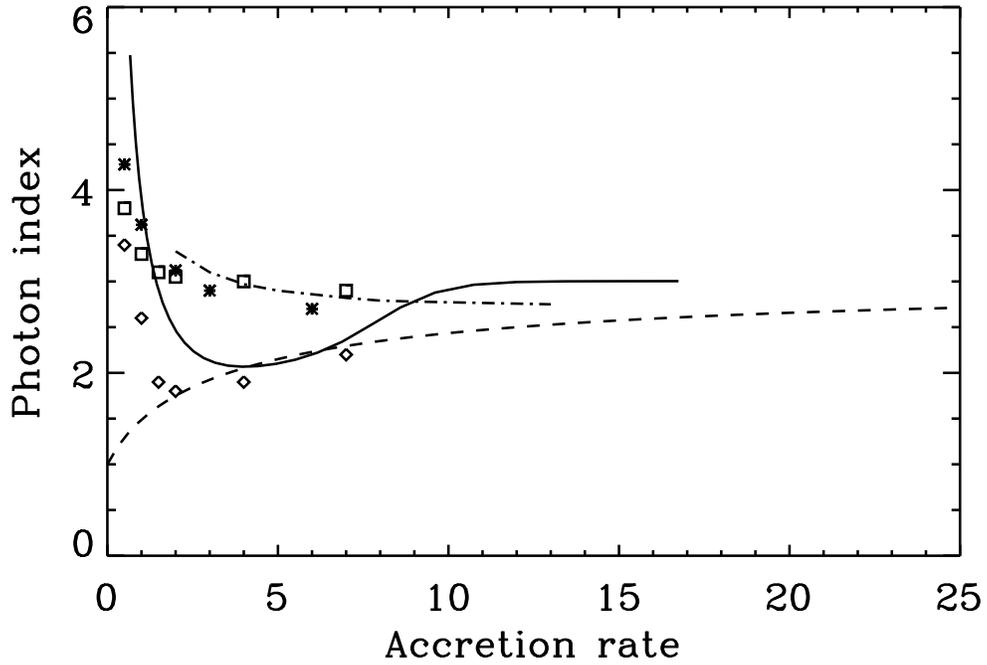}
\caption{\label{compar} The photon index vs. $\dot m$ as derived from
different studies: non-relativistic diffusion approximation (present
investigation, full line), Titarchuk \& Zannias (dash-dotted line),
relativistic diffusion approximation (Turolla et al., dashed line),
Papathanassiou \& Psaltis (crosses), Laurent \& Titarchuk (Schwarzschild
geometry, open squares; flat geometry, diamonds
 }
\end{figure}

\end{document}